\documentclass{article}
\usepackage{amssymb}
\usepackage{amsfonts}

\begin{document}

\title{Complex hidden symmetries in real spacetime and their algebraic
structures}
\author{R. Vilela Mendes\thanks{%
rvilela.mendes@gmail.com; rvmendes@ciencias.ulisboa.pt;
https://label2.tecnico.ulisboa.pt/vilela/} \\
CMAFcIO, Faculdade de Ci\^{e}ncias, Univ. Lisboa}
\date{ }
\maketitle

\begin{abstract}
Considering real spacetime as a Lorentzian fiber in a complex manifold,
there is a mismatch of the elementary linear representations of their
symmetry groups, the real and complex Poincar\'{e} groups. No spinors are
allowed as linear irreducible representations for the complex case, but when
a spin$^{h}$ structure is implemented on the associated principal bundles,
one is naturally led to an algebraic structure similar to the one of the
standard model. This last (dynamical) structure might therefore be inherited
from the kinematical symmetries of a larger space.
\end{abstract}

\section{Introduction}

The coordinates, used in the space-time we live in, take values on the
reals, with four dimensions and a Lorentzian metric. However in many
instances, in the formulation of physical theories, it is convenient to
consider the real field to be embedded into higher dimension division
algebras. For example, in quantum field theory calculations it is convenient
to study Wick-rotated Green's functions, and complex extensions shed light
on the study of singularities and topological effects in general relativity 
\cite{Einstein} \cite{Kunstatter} \cite{Plebanski} \cite{Esposito}. Also, in
the framework of complex manifolds, twistor theory \cite{Penrose} renders
null lines and null surfaces as the basic building blocks, with space-time
points as derived entities, etc. A more extreme view, of course, would be to
consider that our real-coordinates space-time is a lower dimensional
subspace of an actual larger space, of which our senses are unaware of.

Whatever the point of view, a relevant question is: When real space-time is
embedded into a higher dimensional division algebra structure, what
properties of the higher structure are inherited or non-inherited by the
real space-time? In particular this refers to the symmetry groups and their
representations. Here, the following particular setting will be considered:
In a complex manifold $\mathcal{M}$ with (local) metric $G=\left(
1,-1,-1,-1\right) $ and symmetry group $L_{\mathbb{C}}\mathcal{=}U\left(
1,3\right) $%
\[
\Lambda ^{\dag }G\Lambda =G
\]%
$\left( \Lambda \in L_{\mathbb{C}}\right) $, real space-time is considered
to be a $4-$dimensional subspace $\mathcal{S}$ with Lorentzian metric, that
is, a Lorentzian fiber in the Grassmannian $Gr_{4}\left( \mathcal{M}\right) $
of four-dimensional frames of $\mathcal{M}$. The symmetry group of $\mathcal{%
S}$ is the Lorentz group $L_{\mathbb{R}}\mathcal{=}SO\left( 1,3\right) $.
Because here only local manifold effects are considered, one also includes
as symmetries of $\mathcal{M}$ and $\mathcal{S}$, respectively, the complex
and real translations. Therefore the (local) symmetry group of $\mathcal{M}$
is the $24-$parameter complex Poincar\'{e} group $P_{\mathbb{C}}$ and the
(local) symmetry group of $\mathcal{S}$ is the $10-$parameter real Poincar%
\'{e} group $P_{\mathbb{R}}$.

According to a well established tradition \cite{Wigner}, the elementary
physical states corresponding to these symmetry groups would be their
unitary irreducible representations. When these were studied \cite%
{Vilela-IJMPA}, a remarkable difference was found between the elementary
representations of $P_{\mathbb{C}}$ and tose of $P_{\mathbb{R}}$. Namely,
half-integer spin states are compatible with $P_{\mathbb{R}}$ but not with $%
P_{\mathbb{C}}$ and, for integer spin states, a superselection rule emerges
when discrete symmetries are included \cite{Vilela-IJMPA} \cite{dark}. Here
the main concern will be the half-integer spin states. Their absence as
elementary states of $P_{\mathbb{C}}$ may be physically interpreted as
meaning that such states cannot be "rotated" away from each fiber to the
other fibers of the Grassmannian $Gr_{4}\left( \mathcal{M}\right) $ of
four-dimensional frames in $\mathcal{M}$. As for symmetries, one might
accept that the symmetry group of each Lorentzian fiber is simply the real
Poincar\'{e} group $P_{\mathbb{R}}$ and that nothing is inherited from the
larger complex group $P_{\mathbb{C}}$. Alternatively one might insist that $%
P_{\mathbb{C}}$ is actually also the symmetry group of the fiber, but that
the symmetry is implemented in a subtler way. This second point of view
leads to an analysis of when spin structures are or are not realized in
coset manifolds. And to what additional quantum numbers appear when one
insists on their implementation.

The coset manifolds that will be considered are those associated to the
isotropy (little) groups of massive and massless states, $SU\left( 3\right)
/SO\left( 3\right) $, $SO\left( 3\right) /SO\left( 2\right) $. The tools of
algebraic topology provide a quick answer to the existence or non-existence
of spin states, but the exact nature of the additional algebraic structures
that appear, when spin structures are implemented, requires a detailed
analysis of the group representations. A partial announcement announcement
of these results was included in a recent letter \cite{Hidden-1}.

The commutation relations of the Lie algebras of $P_{\mathbb{C}}$ and $P_{%
\mathbb{R}}$ are listed in Appendix A, with $p_{\mathbb{C}}=\left\{ K_{\mu
},H_{\mu },M_{\mu \nu },N_{\mu \nu }\right\} $ and $p_{\mathbb{R}}=\left\{
K_{\mu },M_{\mu \nu }\right\} $. A convenient $4-$dimensional matrix
representation of $M_{\mu \nu }$ and $N_{\mu \nu }$ is included in Appendix
B.

\section{The coset manifold SU(3)/SO(3)}

The elementary unitary representations of $P_{\mathbb{C}}$ are classified by
the eigenvalues of the first Casimir operator $P^{2}$%
\begin{equation}
P^{2}=K^{\mu }K_{\mu }+H^{\mu }H_{\mu }  \label{CM1.1}
\end{equation}%
and the representations of their isotropy (little) groups. For massive
states, $P^{2}>0$, the little group is $U\left( 3\right) \simeq U\left(
1\right) $ $\oplus SU\left( 3\right) $ with Lie algebra (cf. Appendix B)

\begin{equation}
\mathcal{L}_{U\left( 3\right) }=\left\{
R_{i},\;U_{i},\;C_{1},C_{2},C_{3}\right\} ;i=1,2,3,  \nonumber
\end{equation}%
the generators for $SU\left( 3\right) $ and $U\left( 1\right) $ being

\begin{eqnarray*}
\mathcal{L}_{SU\left( 3\right) } &=&\left\{
R_{i},\;U_{i},\;C_{1}-C_{2},C_{1}+C_{2}-2C_{3}\right\} ;i=1,2,3 \\
\mathcal{L}_{U\left( 1\right) } &=&\left\{ C_{1}+C_{2}+C_{3}\right\}
\end{eqnarray*}

For $P_{\mathbb{R}}$ the first Casimir operator is%
\begin{equation}
p^{2}=K^{\mu }K_{\mu }  \label{CM1.3}
\end{equation}%
The rotation subgroup in $SU\left( 3\right) $ is generated by $\left\{
R_{i}\right\} $, which are also the generators of the little group for
massive states in the real fiber $\mathcal{S}$%
\begin{equation}
\mathcal{L}_{SO\left( 3\right) }=\left\{ R_{i}\right\}   \label{CM1.4}
\end{equation}%
The little group $SO\left( 3\right) $ in each real fiber has (double-valued)
half-integer spin representations, whereas $SU\left( 3\right) $ has no
half-integer representations for the subgroup generated by $\left\{
R_{i}\right\} $. This has a clear algebraic topology interpretation when the 
$SU\left( 3\right) $ group is represented as a $SO\left( 3\right) $
principal bundle $\mathcal{P}_{SO\left( 3\right) }$. The base of this bundle
is the coset manifold $SU\left( 3\right) /SO\left( 3\right) $ and existence
of half-integer states in $SU\left( 3\right) $ is synonymous with the
problem of existence of spin structures in the coset manifold. A spin
structure implies the existence of a double cover $\mathcal{P}_{Spin\left(
3\right) }$ of $\mathcal{P}_{SO\left( 3\right) }$ and a commutative diagram%
\[
\begin{array}{c}
\mathcal{P}_{Spin\left( 3\right) }\rightarrow \mathcal{P}_{SO\left( 3\right)
} \\ 
\searrow \;\swarrow  \\ 
SU\left( 3\right) /SO\left( 3\right) 
\end{array}%
\]%
A necessary and sufficient condition for the existence of a spin structure
on the bundle is the vanishing of the second Stiefel-Whitney class \cite%
{spingeo}. Let%
\begin{equation}
\mathcal{L}_{SU\left( 3\right) }=\mathcal{H}^{(1)}\mathcal{\oplus K}^{(1)}%
\mathcal{=}\left\{ R_{i}\right\} \oplus \left\{
U_{i},C_{1}-C_{2},C_{1}+C_{2}-2C_{3}\right\}   \label{CM1.5}
\end{equation}%
Because $\left[ \mathcal{H}^{(1)},\mathcal{K}^{(1)}\right] \subset \mathcal{K%
}^{(1)}$ and $\left[ \mathcal{K}^{(1)},\mathcal{K}^{(1)}\right] \subset 
\mathcal{H}^{(1)}$ the coset space $SU\left( 3\right) /SO\left( 3\right) $
is reductive and a symmetric space. $SU\left( 3\right) /SO\left( 3\right) $
is a $5-$dimensional manifold known as the Wu manifold \cite{Wu} \cite{Atlas}
\cite{Debray}. Its topological properties are known. Its Stiefel-Whitney
class being nonzero, it is not a spin-manifold, nor a spin$^{c}$-manifold.
It is however a spin$^{h}$-manifold \cite{Chen-Th} \cite{spinh}, meaning
that to implement a spin structure over $SU\left( 3\right) /SO\left(
3\right) $ one has to add, to the $\mathcal{P}_{SO\left( 3\right) }$
manifold, another manifold $\mathcal{P}_{SU\left( 2\right) }^{\prime }$ with
a quaternion ($Sp\left( 1\right) \simeq SU\left( 2\right) $) structure group
in the fibers and the same base space (a Whitney sum $\mathcal{P}_{SO\left(
3\right) }\oplus \mathcal{P}_{SU\left( 2\right) }^{\prime }$). In this way
one may have spinor states, consistent with both the $U\left( 3\right) $ of
the full isotropy group of massive states and the $SO\left( 3\right) $ of
the fibers. Hence, the physical consequences of insisting on a full $P_{%
\mathbb{C}}-$ symmetry for spinor massive states would be:

- The emergence of additional quantum numbers associated to the $\mathcal{P}%
_{SU\left( 2\right) }^{\prime }$ auxiliary bundle..

- A degeneracy of the spinor states parametrized by the coset space $%
SU(3)/SO(3)\footnote{%
actually $U(3)/SO(3).$}$, where a $SU\left( 3\right) $ group operates
transitively.

This is as far as one may go from the algebraic topology data. More precise
information on the nature of the additional quantum numbers is obtained from
comparing the representations of $SU\left( 3\right) $ with those in $%
\mathcal{P}_{SO\left( 3\right) }\oplus \mathcal{P}_{SU\left( 2\right)
}^{\prime }$.

The following conclusions might probably be obtained from simple inspection
of the matrices in the Appendix B, but become clearer by using a creation
operator representation of the Lie algebra of $SU\left( 3\right) $. Let $%
\left\{ a_{i}^{\dag },a_{i};i=1,2,3\right\} $ be a set of bosonic creation
and annihilation operators. Then the generators of the $SU\left( 3\right) $
algebra have the representation%
\begin{eqnarray}
R_{i} &=&\epsilon _{ijk}a_{j}^{\dag }a_{k};\;U_{i}=i\left\vert \epsilon
_{ijk}\right\vert a_{j}^{\dag }a_{k};\;C_{1}-C_{2}=i\sqrt{2}\left(
a_{1}^{\dag }a_{1}-a_{2}^{\dag }a_{2}\right) ;\;  \nonumber \\
C_{1}+C_{2}-2C_{3} &=&i\sqrt{2}\left( a_{1}^{\dag }a_{1}+a_{2}^{\dag
}a_{2}-2a_{3}^{\dag }a_{3}\right)   \label{CM1.6}
\end{eqnarray}%
A $SU\left( 3\right) $ triplet is formed by $3$ spin-one states%
\begin{equation}
\left\vert 1\right) =\frac{1}{\sqrt{2}}\left( a_{1}^{\dag }+ia_{2}^{\dag
}\right) \left\vert 0\right\rangle ;\;\left\vert -1\right) =\frac{1}{\sqrt{2}%
}\left( a_{1}^{\dag }-ia_{2}^{\dag }\right) \left\vert 0\right\rangle
;\;\left\vert 0\right) =a_{3}^{\dag }\left\vert 0\right\rangle 
\label{CM1.7}
\end{equation}%
Now one has to match these states by composing an observable spin $\frac{1}{2%
}$ state in the real fiber with another spin $\frac{1}{2}$ state from the
auxiliary $\mathcal{P}_{SU\left( 2\right) }^{\prime }$ bundle. Notice
however that the three states in (\ref{CM1.7}) not only transform under the $%
R_{i}$ rotation generators, but also have nontrivial transformation
properties under the other $SU\left( 3\right) $ generators, in particular
the $U_{i}^{\prime }$s. This must \ also be matched by the composite states.

Let $c_{i}^{\uparrow \dag }$ and $c_{j}^{\downarrow \dag }$ be the spin up
and spin down creator operators of the observable state in the real fiber
and $b_{i}^{\uparrow \dag }$ and $b_{j}^{\downarrow \dag }$ the
corresponding operators of the auxiliary bundle. The $i$ and $j$ labels
stand for the additional quantum numbers needed to match the $SU\left(
3\right) $ transformation properties of the states. The following scheme
satisfies the matching requirements%
\begin{equation}
\left\vert 1\right) \rightarrow b_{i}^{\uparrow \dag }c_{i}^{\uparrow \dag
}\left\vert 0\right\rangle ;\;\left\vert -1\right) \rightarrow
b_{j}^{\downarrow \dag }c_{j}^{\downarrow \dag }\left\vert 0\right\rangle
;\;\left\vert 0\right) \rightarrow \left( b_{i}^{\uparrow \dag
}c_{j}^{\downarrow \dag }+b_{j}^{\downarrow \dag }c_{i}^{\uparrow \dag
}\right) \left\vert 0\right\rangle   \label{CM1.8}
\end{equation}%
Notice that the "internal" quantum numbers $i$ and $j$ of the spinor states $%
c_{i}^{\uparrow \dag }$ and $c_{j}^{\downarrow \dag }$ may be different for
different spin projections, showing the chiral nature of this quantum
number. The spinor doublets in the real fiber are thus endowed with a
doublet of "internal" quantum numbers, that is, a $SU\left( 2\right) $
structure inherited from the $\mathcal{P}_{SU\left( 2\right) }^{\prime }$
bundle. On the other hand when two such spinors are coupled as in (\ref%
{CM1.8}) they form a state that transforms under $SU\left( 3\right) $.
Therefore the spin$^{h}$ mechanism not only generates an additional $%
SU\left( 2\right) $ structure on the elementary states, but also a kind of
"flavor $SU\left( 3\right) $" on the coupled states. This $SU\left( 3\right) 
$ structure, that at the complex space-time level would be an exact
symmetry, may connect states with different $p^{2}=K^{\mu }K_{\mu }$values
(the real mass square) because they are eigenvalues of $P^{2}$, not of $p^{2}
$, and from the commutation relations in Appendix A one sees how the $U_{i}$
operators permute the $K_{\mu }$ and $H_{\mu }$ operators. Summarizing, an
exact complex symmetry may appear as a broken symmetry in the real fibers.
An additional feature of the (\ref{CM1.8}) structure is the existence of a
spin zero state%
\[
\left( b_{i}^{\uparrow \dag }c_{j}^{\downarrow \dag }-b_{j}^{\downarrow \dag
}c_{i}^{\uparrow \dag }\right) \left\vert 0\right\rangle 
\]

With the additional $\mathcal{P}_{SU\left( 2\right) }^{\prime }$ bundle,
spinor states are consistently constructed at each point of the coset space $%
SU(3)/SO(3)$. Therefore, a second consequence is a degeneracy of the spinor
states parametrized by the points of the coset space. Recalling \cite%
{Vilela-IJMPA} that the little group of massive states for the complex
Poincar\'{e} group $P_{\mathbb{C}}$ is actually $U\left( 3\right) $ one
should in fact consider the $6-$dimensional manifold $\mathcal{M}%
^{(1)}=U(3)/SO(3)$. A detailed characterization of this space is important,
particularly if the $U(3)$ symmetry, in its double role, is gauged. A
coordinate system in $\mathcal{M}^{(1)}$ is obtained by choosing a point in
each coset $\left( \mathbf{x}\right) $ and when an element $g\in U(3)$ acts
upon $\left( \mathbf{x}\right) $%
\begin{equation}
g\left( \mathbf{x}\right) =\left( \mathbf{x}^{\prime }\right) h\left( 
\mathbf{x},g\right)   \label{CM1.9}
\end{equation}%
with $h\left( \mathbf{x},g\right) \in SO\left( 3\right) $.

In the spinor degeneracy manifold $\mathcal{M}^{(1)}=U\left( 3\right)
/SO\left( 3\right) $, which will be tentatively called the "color manifold",
a set of coordinates may be obtained in several ways:

1) Let $A=\left\{ a_{ij}\right\} $ be an arbitrary $U\left( 3\right) $
matrix. Then $\Sigma _{j=1}^{3}a_{ij}a_{ij}^{\ast }=1$ but $\Sigma
_{j=1}^{3}a_{ij}a_{ij}=C_{i}$, a non trivial complex number. Acting on $A$
on the right with a $SO\left( 3\right) $ matrix, $C_{i}$ remains invariant.
Therefore the $3$ complex numbers $C_{i}$ $\left( i=1,2,3\right) $ provide a
set of $6$ coordinates for $U(3)/SO(3)$.

2) Alternatively, starting from the identity of $U(3)$, $U(3)/SO(3)$ may be
parametrized by choosing a representative element in each coset. Choose a
symmetric set of coordinates in $\mathcal{M}^{(1)}$ by defining%
\begin{eqnarray}
Q_{1} &=&U_{1}+C_{1};Q_{2}=U_{1}-C_{1}  \nonumber \\
Q_{3} &=&U_{2}+C_{2};Q_{4}=U_{2}-C_{2}  \nonumber \\
Q_{5} &=&U_{3}+C_{3};Q_{6}=U_{3}-C_{3}  \label{CM1.10}
\end{eqnarray}%
with the representative group element in each coset being%
\begin{equation}
\left( \mathbf{x}\right) =e^{\left( x_{1}Q_{1}+x_{2}Q_{2}\right) }e^{\left(
x_{3}Q_{3}+x_{4}Q_{4}\right) }e^{\left( x_{5}Q_{5}+x_{6}Q_{6}\right) }
\label{CM1.11}
\end{equation}%
Using the matrices in Appendix B, $\left( \mathbf{x}\right) $ is represented
by the $3\times 3$ matrix $L\left( \mathbf{x}\right) $ with elements%
\begin{eqnarray}
L_{11}\left( \mathbf{x}\right) &=&e^{i\sqrt{2}(x_{1}-x_{2})}\cos \left(
x_{3}+x_{4}\right) \cos \left( x_{5}+x_{6}\right)  \nonumber \\
L_{12}\left( \mathbf{x}\right) &=&e^{i\sqrt{2}(x_{1}-x_{2})}\cos \left(
x_{3}+x_{4}\right) \sin \left( x_{5}+x_{6}\right)  \nonumber \\
L_{13}\left( \mathbf{x}\right) &=&ie^{i\sqrt{2}(x_{1}-x_{2})}e^{i\sqrt{2}%
(x_{5}-x_{6})}\sin \left( x_{3}+x_{4}\right)  \nonumber \\
L_{21}\left( \mathbf{x}\right) &=&-\sin \left( x_{1}+x_{2}\right) \sin
\left( x_{3}+x_{4}\right) \cos (x_{5}+x_{6})  \nonumber \\
&&+ie^{i\sqrt{2}(x_{3}-x_{4})}\cos \left( x_{1}+x_{2}\right) \sin \left(
x_{5}+x_{6}\right)  \nonumber \\
L_{22}\left( \mathbf{x}\right) &=&-i\sin \left( x_{1}+x_{2}\right) \sin
\left( x_{3}+x_{4}\right) \sin (x_{5}+x_{6})  \nonumber \\
&&+e^{i\sqrt{2}(x_{3}-x_{4})}\cos \left( x_{1}+x_{2}\right) \cos \left(
x_{5}+x_{6}\right)  \nonumber \\
L_{23}\left( \mathbf{x}\right) &=&ie^{i\sqrt{2}(x_{5}-x_{6})}\sin \left(
x_{1}+x_{2}\right) \cos \left( x_{3}+x_{4}\right)  \nonumber \\
L_{31}\left( \mathbf{x}\right) &=&i\cos \left( x_{1}+x_{2}\right) \sin
\left( x_{3}+x_{4}\right) \cos (x_{5}+x_{6})  \nonumber \\
&&-e^{i\sqrt{2}(x_{3}-x_{4})}\sin \left( x_{1}+x_{2}\right) \sin \left(
x_{5}+x_{6}\right)  \nonumber \\
L_{32}\left( \mathbf{x}\right) &=&-i\sin \left( x_{1}+x_{2}\right) \sin
\left( x_{3}+x_{4}\right) \sin (x_{5}+x_{6})  \nonumber \\
&&+e^{i\sqrt{2}(x_{3}-x_{4})}\cos \left( x_{1}+x_{2}\right) \cos \left(
x_{5}+x_{6}\right)  \nonumber \\
L_{33}\left( \mathbf{x}\right) &=&e^{i\sqrt{2}(x_{5}-x_{6})}\cos \left(
x_{1}+x_{2}\right) \cos \left( x_{3}+x_{4}\right)  \label{CM1.12}
\end{eqnarray}

The set $\left\{ Q_{i};i=1,\cdots ,6\right\} $ is an orthogonal set%
\[
Tr\left( Q_{i}Q_{j}\right) =-2\delta _{ij}
\]%
A (Riemannian) metric $g$ may be established in $\mathcal{M}^{(1)}$ by
transporting this quadratic form by the group elements $L\left( \mathbf{x}%
\right) $%
\[
g_{\mathbf{x}}\left( X,Y\right) =Tr(L^{-1}\left( \mathbf{x}\right) ^{\ast
}XL^{-1}\left( \mathbf{x}\right) ^{\ast }Y)
\]%
with $X,Y\in T\mathcal{M}^{(1)}$, the tangent space to $\mathcal{M}^{(1)}$.
The metric is computed from (\ref{CM1.12}) by%
\begin{eqnarray}
g_{ij} &=&Tr\left( L^{-1}\left( \mathbf{x}\right) \left( \partial
_{x_{i}}L\left( \mathbf{x}\right) \right) L\left( \mathbf{x}\right)
L^{-1}\left( \mathbf{x}\right) \left( \partial _{x_{j}}L\left( \mathbf{x}%
\right) \right) L\left( \mathbf{x}\right) \right)   \nonumber \\
&=&Tr\left( \partial _{x_{i}}L\left( \mathbf{x}\right) \partial
_{x_{j}}L\left( \mathbf{x}\right) \right)   \label{CM1.13}
\end{eqnarray}%
From the L\'{e}vy connection associated to this metric a set of gauge fields
is obtained in the "color manifold" $\mathcal{M}^{(1)}$. Notice however that
because of the Whitney sum $\mathcal{P}_{SO\left( 3\right) }\oplus \mathcal{P%
}_{SU\left( 2\right) }^{\prime }$ there is also a connection for the $%
SU\left( 2\right) $ structure and, therefore, new gauge fields associated to
the degrees of freedom associated to the $SU\left( 2\right) $ structure.

The metric so obtained in the $\mathbf{x}$ coordinate representation has
complex nondiagonal elements for arbitrary points of the $\mathcal{M}^{(1)}$
manifold. A diagonal metric metric may be obtained on a vielbein of
Lie-algebra valued $1-$forms $\Gamma _{k}$ by%
\begin{eqnarray*}
\Omega \left( \mathbf{x}\right)  &=&L^{-1}\left( \mathbf{x}\right) \partial
_{x_{i}}L\left( \mathbf{x}\right) dx_{i} \\
\Gamma _{k}\left( \mathbf{x}\right)  &=&Tr\left( Q_{k}\Omega \left( \mathbf{x%
}\right) \right) 
\end{eqnarray*}%
$\Gamma _{k}^{2}$ are then the diagonal elements of the metric, but the
complexity is, of course, in the vielbein.

Summarizing: When $P_{\mathbb{C}}-$ symmetry is inherited by the real
Lorentzian fibers, existence of spinor states implies a degeneracy of
massive spinor states, parametrized by a $6-$dimensional manifold as well
as, through a spin$^{h}$ implementation on a Wu manifold, additional degrees
of freedom ($SU\left( 2\right) $ and $SU\left( 3\right) $), which would
correspond to an exact symmetry in the (complex) $P_{\mathbb{C}}$ group, but
may appear as a broken symmetry for $P_{\mathbb{R}}$.

This structure is evocative of the structure of the particle physics
standard model, but is different. In the standard model the color space is a
linear representation of $SU\left( 3\right) $ and the color states are
linear elements of this representation. Here, in contrast, color and
anti-color states would be particular directions in a curved manifold. More
precisely, a frame of sections in a vector bundle. Also in the $SU\left(
3\right) \times SU\left( 2\right) \times U\left( 1\right) $ group of the
standard model, the $SU\left( 2\right) $ structure is an independent
structure, whereas here it appears as a result of the Whitney sum with $%
\mathcal{P}_{SU\left( 2\right) }^{\prime }$ and, when the representations
for composite states are matched, it turns out to be related to the $%
SU\left( 3\right) $ subgroup of the $U\left( 1,3\right) $ group.

\section{Massless states and U(2)/SO(2)}

For the irreducible representations of $P_{\mathbb{C}}$, in the case $%
P^{2}=K^{\mu }K_{\mu }+H^{\mu }H_{\mu }=0,$ $p^{\mu }\neq 0$, the algebra of
the little group $G_{2}^{\mathbb{C}}$ for the momentum $\left(
p^{0},0,0,p^{0}\right) $ is a semi-direct sum of two subalgebras $N^{c}$ and 
$H^{c}$ with $\left[ N^{c},N^{c}\right] \subset N^{c};\;\left[ H^{c},N^{c}%
\right] \subset N^{c};\;\left[ H^{c},H^{c}\right] \subset H^{c}$. $N^{c}$,
the algebra of the normal subgroup $\mathcal{N}^{c}$, is a 2-dimensional
Heisenberg algebra%
\[
N^{c}=\left\{
l_{1}=L_{1}+R_{2};m_{1}=M_{1}+U_{2};l_{2}=L_{2}-R_{1};m_{2}=M_{2}+U_{1};m_{3}=M_{3}+%
\frac{1}{\sqrt{2}}\left( C_{3}-C_{0}\right) \right\} 
\]%
and $H^{c}$ is the algebra of $U\left( 2\right) $%
\[
H^{c}=\left\{ \frac{R_{3}}{2};\frac{U_{3}}{2};\frac{1}{2\sqrt{2}}\left(
C_{1}-C_{2}\right) ;\frac{1}{\sqrt{2}}\left( C_{1}+C_{2}\right) \right\} 
\]%
Notice the normalization of the generators, which is the one that is
consistent with the usual $SU\left( 2\right) $ spectrum of integer and
half-integer values. The representations are obtained by the induced
representation method \cite{Vilela-IJMPA}. As in the $P_{\mathbb{R}}$ case
(the real Lorentzian fibers) there are two types of representations, the
continuous-spin and the discrete-spin representations. Restricting our
attention to this last case, it corresponds to the trivial representations
of $\mathcal{N}^{c}$, that is, the generators $l_{i},m_{i}$ are mapped on
the zero operator. Then, from $\mathcal{H}^{c}=U\left( 2\right) $ one
concludes that the states are labelled by the quantum numbers of $SU\left(
2\right) $ and a $U\left( 1\right) $ phase. Each spin projection in the $%
SU\left( 2\right) $ multiplet may correspond to a different particle state
in the real Lorentzian fiber, because only $R_{3}$ among the $SU\left(
2\right) $ generators belongs to the real Lorentz group.

The coset space in this case is $U\left( 2\right) /SO\left( 2\right) $. $%
SO\left( 3\right) /SO\left( 2\right) \simeq \mathbb{C}P^{1}$ is a spin
manifold, therefore, in principle, there should be no topological
obstruction to the implementation of spinors in the real Lorentz fibers. A
finer analysis, though, is obtained when the global representations of $%
SO\left( 3\right) $ are compared with those of the $SO\left( 2\right) $
fiber. The spectrum of a representation of the group $\mathcal{H}^{c}$ means
that $\frac{R_{3}}{2}$ may take integer or half-integer values. But then $%
R_{3}$ will only take integer values. Therefore to have consistency, $R_{3}=%
\frac{1}{2}$ states in the fiber can only exist if that spin is complemented
by an additional spin $\frac{1}{2}$. The situation, then becomes identical
to the massive case, only now the additional quantum number has to match the 
$SU\left( 2\right) $ structure of $\mathcal{H}^{c}$ rather than the $%
SU\left( 3\right) $ structure of the little group as in the massive case.
Hence one also has the emergence of a "flavor-type" structure in the
massless case, following from the insistence on the (hidden) implementation
of $P_{\mathbb{C}}$ symmetry.

The $SO\left( 3\right) /SO\left( 2\right) $ coset space has two dimensions
which may simply be assigned to the particle-antiparticle duality. No
"color-type" degeneracy appears in the massless case.

\section{Remarks}

1) In the $U\left( 1,3\right) $ context, massive and massless particles
appeared associated to the coset manifolds $U\left( 3\right) /SO\left(
3\right) $ and $SO\left( 3\right) /SO\left( 2\right) $. Another relevant
coset manifold is $SU\left( 1,3\right) /SO\left( 1,3\right) $ which relates
Lorentz invariance with the generalized complex Lorentz invariance. The
Dirac operator, acting in elementary or coupled spinors, has its domain in
vector bundles with basis in this manifold and representations of $SO\left(
1,3\right) $ as fibers.\ Because maximal compact subgroups are homotopical
equivalent to the Lie groups that contain them, one expects \cite{Stasheff}
the same obstructions to a spin structure over $SU\left( 1,3\right)
/SO\left( 1,3\right) $ as in the Wu manifold, $SU\left( 3\right) /SO\left(
3\right) $. A similar conclusion might be obtained from the representations
of the total group $SU\left( 1,3\right) $. The representations of $SU\left(
1,3\right) $ may be decomposed as direct sums of representations of $%
SU\left( 3\right) \otimes U\left( 1\right) $. Then, considering the subgroup
chain%
\[
SU\left( 1,3\right) \supset SU\left( 3\right) \supset SO\left( 3\right) 
\]%
one concludes that there are no half-integer spins associated to the
(physical) $SO\left( 3\right) $ on this chain. As in the Wu manifold, a spin$%
^{h}$ construction is needed to have $4$-component spinors in the fibers.

Several authors in the past, looking for spectrum generating algebras, have
dismissed the $SU(1,3)$ algebra, as useless, for not having irreducible
spinor representations. However, the fact that $SO(1,3)$ has spinor
representations and $SU(1,3)$ does not, is the most interesting feature of
the latter.

2) The basic idea in this paper was the exploration of higher space-time
symmetries, which although not readily apparent in the analysis of linear
representation of the groups, might still be present and manifested in
subtler nonlinear ways. The inspiration was, of course, the embedding of
real space-time in a wider complex domain. However there are other
possibilities. An amusing connection arises from the observation of the
action of the Nambu relativistic string \cite{Nambu}. As already pointed out
by Mita \cite{Mita}, when the string is covariantly quantized in the
orthonormal gauge, the invariance group of the Hamiltonian is exactly $%
SU\left( 1,3\right) $. The possible relevance of this higher group might
therefore have a dynamical origin, instead of being associated to the
kinematical symmetries of a complex space-time. Conversely, one might
instead say that the relevance of the string interpretation of some hadronic
phenomena might have its origin on the complex embedding. The reasoning goes
both ways.

3) Here, the inspiration was the embedding of space-time as a real fiber in
a complex manifold. A relevant question is what would be the consequences of
considering real space time as a real fiber of a space time parametrized by
higher division algebras, quaternions or octonions \cite{Vilela-IJMPA}. The
obstructions to spin and superselection rules in discrete symmetries are
similar to the complex space. However much more quantum numbers are involved
and more hidden symmetries might be at play.

4) In the construction of the representations associated to $SU\left(
1,3\right) $, the translation part has been considered to be the one of the
complex Poincar\'{e} group. That is, the present analysis is a local one,
which does not consider gravitational curvature effects. Likewise possible
small distance non-commutative space-time effects are not considered \cite%
{Vilela-fund-time} \cite{Vilela-Crab}. What would be the effect of embedding
higher hidden symmetries in those contexts is, of course, an interesting
issue to be explored in the future.

\section{Appendix A: Complex Poincar\'{e} group algebra}

The generators of the algebra of the complex Poincar\'{e} group $P_{\mathbb{C%
}}$ are $\left\{ K_{\mu },H_{\mu },M_{\mu \nu },N_{\mu \nu }\right\} $, with 
$K_{\mu }$ and $H_{\mu }$ the generators of real and imaginary translations, 
$M_{\mu \nu }=-M_{\nu \mu }$ real rotations and boosts and $N_{\mu \nu
}=N_{\nu \mu }$ imaginary transformations. The algebra of the real Poincar%
\'{e} group $P_{\mathbb{R}}$ is $\left\{ K_{\mu },M_{\mu \nu }\right\} $. 
\begin{eqnarray*}
\left[ M_{\mu \nu },M_{\rho \sigma }\right] &=&-M_{\mu \sigma }g_{\nu \rho
}-M_{\nu \rho }g_{\mu \sigma }+M_{\nu \sigma }g_{\rho \mu }+M_{\mu \rho
}g_{\nu \sigma } \\
\left[ M_{\mu \nu },N_{\rho \sigma }\right] &=&-N_{\mu \sigma }g_{\nu \rho
}+N_{\nu \rho }g_{\mu \sigma }+N_{\nu \sigma }g_{\rho \mu }-N_{\mu \rho
}g_{\nu \sigma } \\
\left[ N_{\mu \nu },N_{\rho \sigma }\right] &=&M_{\mu \sigma }g_{\nu \rho
}+M_{\nu \rho }g_{\mu \sigma }+M_{\nu \sigma }g_{\rho \mu }+M_{\mu \rho
}g_{\nu \sigma } \\
\left[ M_{\mu \nu },K_{\rho }\right] &=&-g_{\nu \rho }K_{\mu }+g_{\mu \rho
}K_{\nu } \\
\left[ M_{\mu \nu },H_{\rho }\right] &=&-g_{\nu \rho }H_{\mu }+g_{\mu \rho
}H_{\nu } \\
\left[ N_{\mu \nu },K_{\rho }\right] &=&-g_{\nu \rho }H_{\mu }-g_{\mu \rho
}H_{\nu } \\
\left[ N_{\mu \nu },H_{\rho }\right] &=&g_{\nu \rho }K_{\mu }+g_{\mu \rho
}K_{\nu } \\
\left[ K_{\mu },H_{\rho }\right] &=&0
\end{eqnarray*}

\section{Appendix B: Matrix representation of the Lie algebra generators of
U(1,3) and U(3)}

For the characterization of the subalgebras of the Lie algebra of$\ U(1,3)$
it is useful to define the set $\left\{
R_{i},L_{i},U_{i},M_{i},C_{i},C_{0}\right\} $ of $4\times 4$ matrix
representations. For $U\left( 3\right) $ the generators are the lower $%
3\times 3$ blocks of $\left\{ R_{i},U_{i},C_{i}\right\} $ 
\[
R_{1}=\left( 
\begin{array}{llll}
0 & 0 & 0 & 0 \\ 
0 & 0 & 0 & 0 \\ 
0 & 0 & 0 & 1 \\ 
0 & 0 & -1 & 0%
\end{array}%
\right) \;R_{2}=\left( 
\begin{array}{llll}
0 & 0 & 0 & 0 \\ 
0 & 0 & 0 & -1 \\ 
0 & 0 & 0 & 0 \\ 
0 & 1 & 0 & 0%
\end{array}%
\right) \;R_{3}=\left( 
\begin{array}{llll}
0 & 0 & 0 & 0 \\ 
0 & 0 & 1 & 0 \\ 
0 & -1 & 0 & 0 \\ 
0 & 0 & 0 & 0%
\end{array}%
\right) 
\]%
\[
L_{1}=\left( 
\begin{array}{llll}
0 & 1 & 0 & 0 \\ 
1 & 0 & 0 & 0 \\ 
0 & 0 & 0 & 0 \\ 
0 & 0 & 0 & 0%
\end{array}%
\right) \;L_{2}=\left( 
\begin{array}{llll}
0 & 0 & 1 & 0 \\ 
0 & 0 & 0 & 0 \\ 
1 & 0 & 0 & 0 \\ 
0 & 0 & 0 & 0%
\end{array}%
\right) \;L_{3}=\left( 
\begin{array}{llll}
0 & 0 & 0 & 1 \\ 
0 & 0 & 0 & 0 \\ 
0 & 0 & 0 & 0 \\ 
1 & 0 & 0 & 0%
\end{array}%
\right) 
\]%
\[
U_{1}=\left( 
\begin{array}{llll}
0 & 0 & 0 & 0 \\ 
0 & 0 & 0 & 0 \\ 
0 & 0 & 0 & i \\ 
0 & 0 & i & 0%
\end{array}%
\right) \;U_{2}=\left( 
\begin{array}{llll}
0 & 0 & 0 & 0 \\ 
0 & 0 & 0 & i \\ 
0 & 0 & 0 & 0 \\ 
0 & i & 0 & 0%
\end{array}%
\right) \;U_{3}=\left( 
\begin{array}{llll}
0 & 0 & 0 & 0 \\ 
0 & 0 & i & 0 \\ 
0 & i & 0 & 0 \\ 
0 & 0 & 0 & 0%
\end{array}%
\right) 
\]%
\[
M_{1}=\left( 
\begin{array}{llll}
0 & i & 0 & 0 \\ 
-i & 0 & 0 & 0 \\ 
0 & 0 & 0 & 0 \\ 
0 & 0 & 0 & 0%
\end{array}%
\right) \;M_{2}=\left( 
\begin{array}{llll}
0 & 0 & i & 0 \\ 
0 & 0 & 0 & 0 \\ 
-i & 0 & 0 & 0 \\ 
0 & 0 & 0 & 0%
\end{array}%
\right) \;M_{3}=\left( 
\begin{array}{llll}
0 & 0 & 0 & i \\ 
0 & 0 & 0 & 0 \\ 
0 & 0 & 0 & 0 \\ 
-i & 0 & 0 & 0%
\end{array}%
\right) 
\]%
\[
C_{1}=\sqrt{2}\left( 
\begin{array}{llll}
0 & 0 & 0 & 0 \\ 
0 & i & 0 & 0 \\ 
0 & 0 & 0 & 0 \\ 
0 & 0 & 0 & 0%
\end{array}%
\right) \;C_{2}=\sqrt{2}\left( 
\begin{array}{llll}
0 & 0 & 0 & 0 \\ 
0 & 0 & 0 & 0 \\ 
0 & 0 & i & 0 \\ 
0 & 0 & 0 & 0%
\end{array}%
\right) \;C_{3}=\sqrt{2}\left( 
\begin{array}{llll}
0 & 0 & 0 & 0 \\ 
0 & 0 & 0 & 0 \\ 
0 & 0 & 0 & 0 \\ 
0 & 0 & 0 & i%
\end{array}%
\right) 
\]%
\[
C_{0}=\sqrt{2}\left( 
\begin{array}{llll}
i & 0 & 0 & 0 \\ 
0 & 0 & 0 & 0 \\ 
0 & 0 & 0 & 0 \\ 
0 & 0 & 0 & 0%
\end{array}%
\right) 
\]%
Their correspondence to the generators in Appendix A is as follows%
\[
R_{i}=\frac{1}{2}\epsilon _{ijk}M_{jk};L_{i}=M_{01};%
M_{i}=N_{0i};U_{i}=N_{jk};C_{\mu +3}=-\frac{1}{\sqrt{2}}%
N_{\mu \mu }g_{\mu \mu } 
\]

\textbf{Acknowledgments}

Partially supported by Funda\c{c}\~{a}o para a Ci\^{e}ncia e a Tecnologia
(FCT), project UIDB/04561/2020: https://doi.org/10.54499/UIDB/04561/2020

\end{document}